\documentclass[twocolumn,showpacs,preprintnumbers,amsmath,amssymb,aps,prb]{revtex4}
\usepackage{graphicx}
\begin{document}

\title{Commensurability, Jamming, and Dynamics For Vortices in Funnel Geometries}   
\author{    
C. J. Olson Reichhardt and C. Reichhardt} 
\affiliation{
Theoretical Division,
Los Alamos National Laboratory, Los Alamos, New Mexico 87545} 

\date{\today}
\begin{abstract}
With advances in 
fabrication technologies it is now possible 
to create precisely controlled
geometries and pinning landscapes for vortex matter in type-II superconductors. 
Here we use numerical simulations to examine vortex states and dynamics in
periodic funnel geometries 
where a drive is applied in the easy flow direction. 
We show that this system exhibits a number of different 
commensurability effects 
when the vortex configurations match to both the periodicity
of the array and the geometry of the funnels. The vortex 
configurations in this system are generally  
different from those observed for single isolated triangular 
superconducting samples due to the
coupling of vortices in adjacent funnels. At certain matching fields, 
peaks in the critical current are absent 
due to the particular vortex configurations that
occur at these fields. We find that the overall depinning force
increases with increasing vortex density as a result of
the enhanced vortex-vortex interactions caused by a
crowding effect at the funnel tips. 
When a system becomes less mobile as a result of increased particle
interactions, it is said to exhibit a jamming behavior.
Under an applied drive we observe 
a series of elastic and plastic vortex flow phases which produce
pronounced features such as jumps or dips in the transport curves.
In all of the flow phases,
only one vortex can pass through the funnel tip
at a time due to the vortex-vortex repulsion forces. 
As a consequence of this constraint, we observe the remarkable result 
that the sum of the vortex velocities at a fixed drive
remains nearly constant with increasing magnetic field $B$ 
rather than increasing linearly. 
This result is similar to the behavior of sand in an hourglass.  
We also show how  
noise fluctuations can be used to distinguish the different flow phases.    
Our results should be readily generalizable 
to other systems of particles flowing in 
periodic funnel geometries, such as colloids or Wigner crystals.  
\end{abstract}
\pacs{74.25.Uv,74.25.Wx}
\maketitle

\vskip2pc

\section{Introduction}
Advanced nanostructuring techniques permit the creation of specific
superconducting structures for
controlling vortex motion in type-II superconductors. 
Experiments and simulations for square and triangular
artificial pinning arrays 
demonstrate that pronounced commensurability effects 
occur when the number of vortices
is an integer multiple of the number of pinning sites,
and that these effects can be observed as peaks in the 
critical current.\cite{Baert,Martin,Commensurate,Commensurate2,Field,VM}  
Above the first matching field, 
multiple vortices may occupy individual pinning sites,\cite{Field} or 
only a portion of the vortices may be captured by the pinning sites
with the remaining vortices located in
interstitial sites where they 
can still be pinned by the repulsive interactions with vortices
at the pinning sites.\cite{Rosseel,Kara}    
Vortex matter in these periodic pinning arrays can also exhibit 
a remarkably rich variety of distinct dynamical 
phases originally predicted in simulations.\cite{Reichhardt,Misko} 
These include 
one-dimensional flow of interstitial vortices between immobile 
vortices in the pinning sites,
disordered or turbulent phases where the number of moving vortices fluctuates
strongly,
and laminar states where the vortex flow can organize along the rows of
pinning sites.
The transitions between these different states 
appear as specific features in the voltage-current curves,
including negative differential conductivity 
where the number of moving vortices or the average 
vortex velocity decreases
with increasing drive. 
In very recent experiments, these dynamical phases were observed in 
both low temperature \cite{AS} and high temperature \cite{Zhil} superconductors 
with periodic pinning arrays.
It is also possible to create periodic pinning arrays that contain intrinsic
asymmetry, such as with
asymmetric thickness modulation,\cite{Barabasi,Lu} 
funnel geometries,\cite{Wambaugh,P,RatchetN} 
composite pinning sites,\cite{Mosh}  
or arrays of triangular traps.\cite{Martin2,R2,Gn,Dinis}
This asymmetry can 
produce a diode effect when the depinning force is higher in one direction, 
and can give a ratchet effect in which a net dc vortex flow occurs upon
application of an ac drive.\cite{Reim}
Reversals of the ratchet flow from the easy asymmetry direction to the 
hard asymmetry direction can occur as a function of magnetic field and
other parameters
\cite{Mosh,Martin2,R2,Lu}
due to various collective interactions of the vortices.   
One of the earliest proposals for a vortex ratchet 
involved a periodic asymmetric channel or funnel,\cite{Wambaugh} and 
such geometries have now been experimentally fabricated.\cite{P,PN,RatchetN} 
The first experiments on this system
verified the existence of ratchet effects, but also detected boundary
effects caused by the disordered injection of vortices
at the edge of the sample.\cite{P}  
The edge effect can be 
overcome by using sample geometries where the vortices flow in 
annular asymmetric channels so that vortices need not 
enter or exit the sample.\cite{RatchetN}
Such a technique may also provide the 
resolution required to investigate individual vortex motion
through the channels.\cite{W}   

In this work we study the dynamics of vortices in a periodic asymmetric 
channel geometry or funnel array and show that
a variety of new types of vortex dynamics and behaviors can arise, 
including a jamming effect and flow patterns
that are organized such that only one vortex at a time passes
through the funnel tip. 
Such a geometry could be realized by etching 
the edge of a single superconducting strip 
into a periodic funnel shape so that the vortices would flow in a 
true single channel. 
Funnel geometries created with periodic channels
in two-dimensional superconducting samples 
should also exhibit many of the same properties we observe; however,
in these systems vortices located 
between the asymmetric channels could be important and
under high drives could depin. The presence of
vortices in the regions between the channels can
be deduced from the onset of hysteresis in the critical current
curves $I_c(H)$ at fields
higher than the fields at which many of the 
ratchet effects are observed.\cite{RatchetN}    
Although our study focuses 
on interacting vortices in type-II superconductors,
we expect that the same dynamics will occur 
for other systems of repulsively interacting
particles in funnel  geometries. These include charged
colloidal assemblies,\cite{Colloid,Peeters} 
magnetic colloids,\cite{Doyle} charged metallic dots,\cite{Dots} 
and classical electron crystals.\cite{Peeters2} 
Additionally, in recent experiments of ion flow through a single
funnel, effects such as negative differential conductivity 
were observed.\cite{Siwy} Our results 
suggest that the same type of clogging effect 
we observe could be occurring in these 
artificial ion channels.  

\section{Jamming and Clogging in Vortex Matter} 

An important feature that differentiates funnel geometries from the other
vortex ratchet geometries is that in the funnel the vortices
are forced to move through a narrow bottleneck. 
At this constricted point, the repulsive vortex-vortex interactions are
very important and
favor the organization of the vortex flow into a pattern that permits only a
single vortex to pass through the bottleneck at a time.
The bottleneck has many 
features in common with granular hopper geometries where grains flow through
a funnel.  In the granular case,  it is known that 
decreasing the width of the hopper aperture can cause
the flow of grains 
to be impeded or jammed. \cite{Pak,J}
Systems that become immobile due to 
particle-particle interactions are often referred to as
jammed.  The physics of jamming 
has attracted growing interest as a way to understand
many types of loose particle assemblies such as grains, colloids, and emulsions
in situations where these systems exhibit a sudden onset 
of resistance to shear, with possible connections to the 
glass transition.\cite{Liu,Drocco}      
In a granular hopper where grains flow 
through a thin funnel, the jamming effect occurs when some grains 
block the motion of other grains.
The interaction between the grains is short ranged and has a sharp cutoff.  
Vortices also experience a 
repulsive interaction; 
however, it is significantly longer in range and 
smoother than the interaction among grains. 
Understanding how  
systems with intermediate range or longer range interactions  
can develop a jammed state is an open question. 

Generally, in a vortex system,
increasing the effective vortex-vortex interaction strength
reduces the effectiveness of the pinning and 
causes the depinning force to decrease. 
If vortex
matter can exhibit a jamming effect, the opposite 
behavior would occur and the 
system would become more immobile with increasing vortex-vortex interaction
strength.
We note that in the peak effect phenomenon, the 
effective pinning force increases with 
increasing vortex density or temperature.
Many explanations of this effect involve
the reduction of 
the effective vortex-vortex interaction 
force with increasing density or temperature
due to changes in the penetration depth or softening 
of the vortex lattice near $H_{c2}$ or $T_{c}$ \cite{Zimanyi2}
in order to match the normal expectation of increased pinning force
caused by decreased vortex interaction strength.  Superficially, however,
the increased pinning force generated by increased vortex density
resembles a jamming effect.
In simulations of vortex systems 
with more vortices than pinning sites,\cite{Fertig} 
gradually increasing the strength of the vortex-vortex interactions
initially increased the depinning force
since the vortices at the pinning sites 
blocked the free vortices from moving, similar to a jamming effect.
On the other hand, when there are more pinning sites than
vortices, the depinning force monotonically decreases
as the vortex-vortex interaction strength
increases.\cite{Zimanyi2} 
This shows that if vortices are to exhibit jamming
behavior, it must arise from the collective interactions 
of the vortices rather than from vortex-pin interactions alone.
In our system we fix the bare vortex-vortex interaction strength
and study the appearance of jamming and clogging effects
due to density-induced changes in the effective 
importance of the vortex-vortex interactions.

\section{Simulation Techniques} 

\begin{figure}
\includegraphics[width=3.5in]{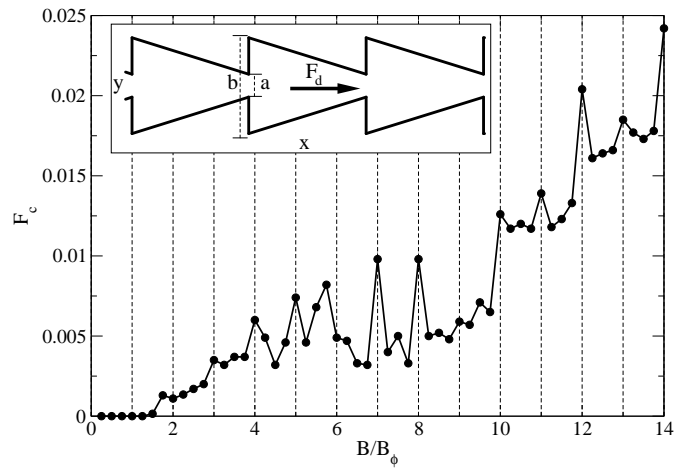}
\caption{
The depinning force $F_c$ vs 
$B/B_{\phi}$ for vortices in a periodic funnel array for driving in
the $+x$, easy-flow direction. 
A series of peaks 
appear at matching fields of 
$B/B_{\phi} = 4$, 5, 7, 8, 11, 12 and 14. Weaker peaks 
occur at $B/B_{\phi} = 3$, 10, and 13, 
while peaks are absent at $B/B_{\phi} = 2$, 6, and 9.      
Inset: A portion of the sample showing the funnel array geometry.
}
\label{fig:1}
\end{figure}

Following previous techniques 
for simulating vortices in periodic pinning geometries,
we employ Langevin dynamics in a two-dimensional 
system.\cite{Brandt,Reichhardt,Misko,Lu,Dinis,Mosh,Wambaugh} 
We consider a sample containing a single channel 
composed of $N_{c}=16$ funnels.
Each funnel has a small aperture size of $a=1.8\lambda$, 
a wide aperture size of
$b=7.4\lambda$, 
and a length 
of $L_c=9\lambda$, where lengths are
measured in units of the London penetration depth $\lambda$.
The funnels are aligned in the $x$ direction and
the sample has periodic boundary conditions along the $x$ direction only. 
A total of $N_v$ 
vortices are placed only inside the funnel channel and the region outside the
channel is empty.
The motion of the vortices is calculated by 
integrating the following overdamped equation of motion:
\begin{equation} 
\eta \frac{d {\bf R}_i}{dt} = -\sum_{i \neq j}^{N_{v}}\nabla U_{vv}(R_{ij}) 
-{\bf F}_{\rm wall}^{i} +
{\bf F}_{D} + {\bf F}^{T}_i . 
\end{equation} 
Here the damping constant is $\eta = \phi_{0}^2d/2\pi \xi^2\rho_N$ in a
crystal of thickness $d$,
where $\phi_{0}=h/2e$ is the elementary flux quantum, $\xi$ 
is the superconducting coherence length,
and $\rho_N$ is the normal state resistivity.
The vortex-vortex force is repulsive with a potential 
$U_{vv}(R_{ij}) = A_{0}K_{0}(R_{ij}/\lambda)$,
where $A_{0} = \phi_{0}^2/2\pi\mu_{0}\lambda^3$,
$K_{0}$ is a modified Bessel function,   
${\bf R}_{i(j)}$ is the position of vortex $i(j)$,
and $R_{ij}=|{\bf R}_i-{\bf R}_j|$.      
If a vortex approaches one of the channel walls sufficiently closely,
it experiences a wall force ${\bf F}_{\rm wall}$.
The channel walls are constructed out of 
$N_b=4N_c$ repulsive elongated potential
barriers which are inverted versions of the potential wells employed in
Ref.~\cite{thermalrca}.
Each barrier has a central rectangular region 
which repels vortices
in the direction transverse to the long direction of the rectangle, along
with two half-parabolic cap regions which repel vortices from the ends
of the barrier.
We have 
\begin{multline}
{\bf F}_{\rm wall}^{i}=(f_p/r_p)
\sum_{k}^{N_b} R_{ik}^\pm \Theta(r_p-R_{ik}^\pm)
\Theta(R_{ik}^\parallel-l_k)
{\bf \hat{R}}_{ik}^\pm \\
+R_{ik}^\perp \Theta(r_p-R_{ik}^\perp)
\Theta(l_k-R_{ik}^\parallel){\bf \hat{R}}_{ik}^\perp .
\end{multline}
Here
$R_{ik}^\pm=|{\bf R}_i-{\bf R}_k^p \pm l_k{\bf {\hat p}}^k_\parallel|$,
$R_{ik}^{\perp,\parallel}=
|({\bf R}_i - {\bf R}_k^p) \cdot {\bf {\hat p}}^k_{\perp,\parallel}|$,
${\bf R}_k^p$
is the position of the center point of barrier $k$, 
$r_p=0.4\lambda$ is the barrier radius or half width,
$f_p=15f_0$ is the barrier strength,
$l_k$ is half the length of the central rectangular
region of barrier $k$,
and ${\bf \hat{p}}^k_\parallel$ (${\bf \hat{p}}^k_\perp$) is a unit
vector parallel (perpendicular) to the axis of barrier $k$.
The individual barriers are connected together 
to form a pair of sawtooth shapes as illustrated in
the inset of Fig.~\ref{fig:1}.  The barrier lengths are 
$2l_k=2.8\lambda$ 
for the
vertical walls and $2l_k=18\lambda/\sqrt{3}$ for the slanted funnel walls.
The central portion of the channel is featureless and the vortices
experience confining forces only from the barrier walls and not from
the channel itself.
The barriers are sufficiently strong that vortices can never cross them
under the conditions considered in this work.

The initial vortex positions are obtained by placing the vortices
evenly throughout the funnel and performing simulated annealing.
Temperature is modeled as Langevin kicks ${\bf F}^{T}$ with the following
properties: $\langle F^{T}(t)\rangle = 0$ and 
$\langle F^{T}_{i}(t)F^{T}_{j}(t^{\prime})\rangle =
2\eta k_{B}T\delta_{ij}\delta(t - t^{\prime})$, 
where $k_{B}$ is the Boltzmann constant. 
After the vortex positions are initialized, we apply an external drive 
${\bf F}_{D}=F_D{\bf {\hat x}}$ 
representing the Lorentz force from an applied current
in the positive $x$ or easy-flow direction 
and measure the average vortex velocity 
$\langle V_{x}\rangle = N_v^{-1}\sum^{N_{v}}_{i} {\bf v}_{i}\cdot {\bf {\hat x}}$. 
The drive is slowly increased in small increments with a fixed waiting
time between each increment.
The waiting time is taken sufficiently long that the system always
reaches a steady state velocity at each drive before we make our
measurements.
The depinning force $F_c$ is defined as the drive at
which $\langle V_x\rangle > 0.001$.
In this work we examine only the dynamics in the
easy flow direction and 
note that in general $F_{c}$  
is higher for driving in the hard or negative $x$ direction so that 
a diode effect is possible, in agreement with experiments.\cite{P,PN} 
For driving in the hard direction, 
clogging type dynamics occur 
which will be explored elsewhere.\cite{Olson2}      

In our system the penetration depth $\lambda$ is less than the funnel size,
so many of our results should carry over to 
typical colloidal systems with optical trap arrays. 
In the vortex ratchet experiments of Ref.~\cite{RatchetN},  
the size of the triangular traps is less than $\lambda$
so we are working in a different regime from the experiment, 
although we expect that many of the same
effects can appear in both systems. 
Experiments with thicker films or larger funnel length scales should also 
be possible which would be much closer to the regime we consider.  

\section{Commensuration Effects in the Easy Flow direction} 

In Fig.~\ref{fig:1} we plot the the depinning force $F_{c}$ for
driving in the easy $+x$ direction vs $B/B_{\phi}$, where $B_{\phi}$ 
denotes the field at which there is one vortex per funnel. 
In our geometry, $F_c=0$ at $B/B_{\phi} = 1.0$ 
since each vortex can move unimpeded through the bottleneck and then flow
freely along the center axis of the funnel.
Fig.~\ref{fig:1} shows peaks in $F_{c}$ at $B/B_{\phi} = 4.0$, 
5.0, 7.0, 8.0, 11.0, 12.0, and 14.0.
Smaller peaks in $F_c$ appear at $B/B_{\phi} = 3.0$, 
10.0, and 13.0, while peaks are absent at 
$B/B_{\phi} = 2.0$, 6.0, and 9.0. 
The varying sizes and shapes of the commensuration peaks are consistent with
the existence of distinct arrangements of singly quantized vortices within
the funnel plaquettes at different matching fields.

Commensurability effects in 
two-dimensional 
square or triangular periodic pinning arrays 
can take two different forms
depending on whether multiple vortices are trapped at each 
pinning site or whether interstitial vortices are present.
If multiple vortex pinning occurs, then the vortex configuration 
at each matching field $B/B_{\phi}=n$, with integer $n$,
is the same as the configuration 
at $B/B_{\phi} = 1.0$, but 
with $n$-quantized vortices trapped at each pinning site.
The result is a peak in the critical current
at every matching field.\cite{Hoffman} 
Similarly, the vortex configurations at 
fractional fields such as $B/B_{\phi} = 1/2$ are repeated 
at all fields $B/B_{\phi}=n + 1/2$.\cite{Field,Kara} 
Such multi-quanta commensuration effects 
are very similar to the commensuration effects observed
in superconducting wire networks.\cite{Itzler} 
If the pinning sites can capture a maximum of one vortex,
then for fields above $B/B_{\phi}=1$,
some vortices will be located in the interstitial regions and the
vortex lattice structures
can be different at each matching field.\cite{Commensurate,Commensurate2} 
At matching fields where the interstitial vortex structure is 
disordered, there is no peak in the
critical current \cite{Commensurate}. 
The magnitude and shape of the commensuration 
peaks show striking variations 
when different types of vortex crystals form at different
matching fields. 
For example, in square pinning 
arrays, a square vortex lattice forms at $B/B_{\phi} = 2.0$, 
a less stable dimer lattice with a smaller critical current peak
appears at $B/B_{\phi} = 3.0$, a very stable triangular lattice with a strong
critical current peak is present at $B/B_{\phi} = 4.0$, 
and the partially disordered vortex structures at $B/B_{\phi} = 6.0$ and $7.0$ 
produce no peaks in the critical current.\cite{Commensurate}

For the asymmetric funnel array, 
the quasi-one-dimensional nature of the system might be expected to produce
identical commensuration effects at each matching field;
however, it is possible for the vortices within each funnel to distort
in both the $x$ and $y$ directions in order to try to form triangular ordering
on a local scale, and thus the response of the system differs at different
matching fields. 
Simulations and experiments 
on single mesoscopic triangular superconducting samples 
have shown that the vortices 
can form triangular or partially
ordered configurations at magic fillings such as 
$B/B_{\phi}= 3.0$, 6.0, and $10.0$.\cite{Zhao} 
At these matching fields, we find weak or missing commensuration 
peaks in the funnel geometry, as indicated in Fig.~\ref{fig:1}. 
The high symmetry of these well ordered states results in poor pinning
of the vortices in the easy-flow direction since it allows some vortices to
simultaneously align along the $x$ axis of the channel while closely 
approaching the narrow aperture of the funnel.  The alignment of the
vortices produces an additional $x$-direction force on the vortex closest
to the funnel tip, permitting it to flow out of the funnel at a relatively
low driving force.
In contrast, at commensurate 
fillings where the vortices adopt partially disordered
configurations or have degenerate ground states, the critical current is
high.  This is likely due to the lack of a well-defined 
easy shear direction for the 
disordered vortices.

Figure \ref{fig:1} shows a vanishing critical current at low 
fields which results when the vortices are able to sit along the center of the
funnel channel and can flow along the channel unimpeded by the funnel
geometry.
The experiments of Ref.~\cite{RatchetN} revealed
a finite critical depinning force at low fields. 
Since the experimental channels are less than $\lambda$ wide, this finite
critical force could be the result of a bowing effect in the center of
the funnel channel that creates some effective pinning even for very low
fields.
The edge barrier may also be playing a role at low fields, and it is
likely that 
some intrinsic random pinning 
exists throughout the entire sample which would create a 
finite $F_{c}$ at all fields. 
If the intrinsic pinning is weak, it should be possible to observe the
commensuration effects shown in Fig.~\ref{fig:1}.
The intrinsic pinning effects 
can be strongly suppressed near $T_c$, and in a later
section we show that the commensuration effects 
in Fig.~\ref{fig:1} are robust at finite temperature.   
We also note that for colloidal particles         
in an optical funnel trap array, 
such intrinsic pinning effects would not be present. 

At incommensurate fields $n<B/B_{\phi}<n+1$, 
due to the discreteness of the vortices the
funnels are occupied by a mixture of $n$ and $n+1$ vortices.
The depinning force is reduced at these fields as a result of the asymmetric
repulsive force experienced by a vortex at the boundary between two
funnels of different occupancy.
Fig.~\ref{fig:1} indicates that peaks or 
enhancements of the depinning force can arise at non-matching fields
such as at $B/B_{\phi} = 3.5$ and $B/B_{\phi}=5.75$; 
however, unlike the matching peaks,
these non-matching peaks are not robust against thermal fluctuations.

\begin{figure}
\includegraphics[width=3.5in]{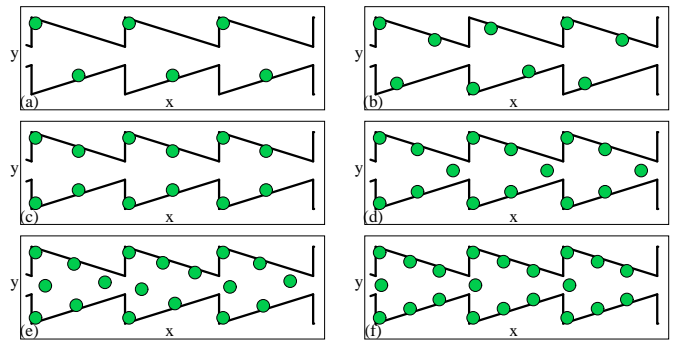}
\caption{
Vortex positions (dots) and funnel geometry (lines) in a small portion
of the sample at different fields with no applied drive.
(a) At $B/B_{\phi} = 2.0$, an aligned dimer state forms.
(b) At $B/B_{\phi}=3.0$, a triangular state forms with the triangle orientation
alternating in every other funnel.
(c) $B/B_{\phi}=4.0$. (d) $B/B_{\phi}=5.0$. 
(e) At $B/B_{\phi}=6.0$, a triangular structure 
forms that fits well with the triangular funnel geometry and that matches
the vortex configuration found in an individual mesoscopic triangular
superconductor. 
This field also
corresponds to a missing commensuration peak in Fig.~\ref{fig:1}. 
(f) $B/B_{\phi}=7.0$.    
}
\label{fig:2}
\end{figure}

In Fig.~\ref{fig:2}(a-f) we illustrate the vortex configurations at zero applied
drive for 
$B/B_{\phi} = 2.0$, 3.0, 4.0, 5.0, 6.0, and 7.0.
For $B/B_{\phi} = 2.0$, 
Fig.~\ref{fig:2}(a) shows that each funnel captures two vortices 
which form a dimer state. 
All of the dimers are aligned in the same direction with
one vortex in each dimer 
located at the upper left corner of the funnel and the 
other vortex located along the lower wall. 
We note that this is not the same configuration that would arise for a
single isolated triangular superconductor, 
where the two vortices would maximize their spacing by sitting with one
vortex at the tip of the triangle and the other vortex in the center of
the opposing triangle wall.
Interactions between vortices in adjacent funnels in our system would make
such an arrangement energetically unfavorable since the vortex at the tip of
the triangle would be too close to the leftmost vortex in the adjacent funnel.
The tilted dimer configuration in Fig.~\ref{fig:2}(a) minimizes the vortex-vortex
interactions both within a single funnel and in adjacent funnels.

The vortex dimer state in Fig.~\ref{fig:2}(a) is two-fold 
degenerate; in the other possible orientation, a vortex is located at the
lower left corner of each funnel.
In an infinitely long system at finite temperature
or in the presence of quenched disorder, 
it is possible that domain wall excitations could form
where the dimer orientation flips from one ground state to the other. 
In this case, it may be possible to map the dimer state
to a one-dimensional Ising model which is 
known to have a long-range ordered ground state 
only at $T = 0.0$. The mapping of
effective dimer and
trimer states of 
particles in periodic substrates to Ising and other spin models has 
been proposed 
previously for colloids on two-dimensional 
periodic substrates \cite{Frey} and vortices
in honeycomb pinning arrays.\cite{Honeycomb} 
For $B/B_{\phi} = 3.0$, Fig.~\ref{fig:2}(b) shows that the three vortices
in each funnel form a triangle with one vortex located in the
corner of the funnel. 
The orientation of the triangle alternates in every other 
plaquette from having
the upper funnel corner occupied by a vortex to having the lower funnel
corner occupied by a vortex.
This ground state has
similarities to a one-dimensional 
antiferromagnetic ordering. 

For $B/B_{\phi} = 4.0$, shown in Fig.~\ref{fig:2}(c), both corners of each funnel are
occupied by vortices and the ground state
is non-degenerate.
This is the first filling at which a pronounced peak in $F_c$ emerges, as
seen in Fig.~\ref{fig:1}.
The ground state
at $B/B_{\phi} = 5.0$,  illustrated in Fig.~\ref{fig:2}(d), is very similar
to the configuration at $B/B_{\phi} = 4.0$ with the addition of
one vortex in the center of the channel near the tip of the funnel, while
at $B/B_{\phi} = 6.0$, 
shown in Fig.~\ref{fig:2}(e), 
there are two vortices in the center of the channel. 
At $B/B_{\phi} = 7.0$, Fig.~\ref{fig:2}(f) indicates that the configuration changes  
from states with four vortices along the walls and the remaining vortices in
the center of the channel to a state with six vortices lining the walls and
only one vortex in the center of the channel.
Figure~\ref{fig:1} indicates that there is a pronounced peak 
in $F_c$ at $B/B_{\phi} = 7.0$ but not at $B/B_{\phi} = 6.0$. 
The vortex configurations at $B/B_{\phi} = 2.0$, 3.0, 4.0, 5.0, and $7.0$
differ from the configurations 
found in isolated mesoscopic triangular superconductors;\cite{Zhao}
however, 
the configuration at $B/B_{\phi} = 6.0$ is almost the same 
as that in an isolated triangle since the vortices can form an almost
perfect triangular ordering within the funnel at this field.
Since the accommodation of the triangular
ordering to the boundaries is so energetically favorable,
it overcomes the energy cost of placing two 
vortices close together near 
the funnel aperture with one vortex shifted
slightly in the positive $x$-direction. 
Since this vortex experiences an extra force
from the other vortices, 
the effective pinning potential at the tip of the funnel for this vortex is
depressed, lowering $F_c$.
At $B/B_{\phi} = 7.0$ this 
condition no longer holds since the extra vortex 
that had been near the funnel tip shifts to a new location along the
wall, where it no longer exerts an extra force on the vortex at the
center of the channel.

\begin{figure}
\includegraphics[width=3.5in]{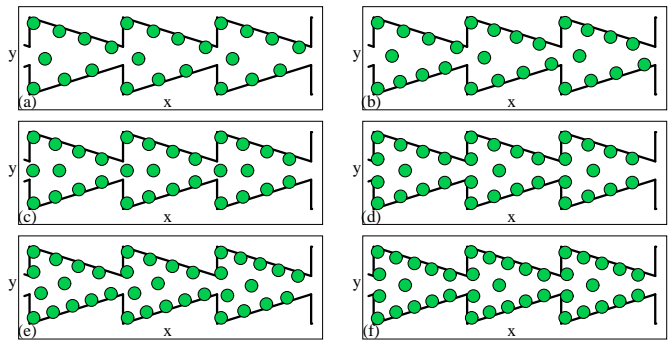}
\caption{
Vortex positions (dots) and funnel geometry (lines) in a small portion of the
sample at different fields with no applied drive.
(a) $B/B_{\phi} = 8.0$. 
(b) $B/B_{\phi}=9.0$, where there is a missing peak in $F_c$. (c) 
$B/B_{\phi}=10.0$.
(d) $B/B_{\phi}=11.0$.  
(e) $B/B_{\phi}=12.0$. (f) 
$B/B_{\phi}=13.0$.  
}
\label{fig:3}
\end{figure}

In Fig.~\ref{fig:3}(a), the ordered vortex configuration at 
$B/B_{\phi} = 8.0$ has four
vortices on the upper wall 
of each funnel and three 
vortices on the bottom wall, along with one vortex 
in the center of the channel near the wide end of the
funnel. This configuration has a two-fold degenerate ground state 
since either the upper or the lower funnel wall 
could be occupied with the four vortices. 
At $B/B_{\phi} = 9.0$, Fig.~\ref{fig:3}(b) 
shows that there are now four vortices lining both the top and bottom
funnel walls.
The vortices on 
one wall are slightly more 
compressed than the vortices on the other wall 
so that the vortices
near the funnel tip are not aligned in the $y$-direction.  The
wall with the stronger compression alternates from top to bottom 
in adjacent funnels. 
Figure~\ref{fig:1} indicates that there is no peak in 
$F_c$ at $B/B_{\phi} = 9.0$.
The relatively low depinning threshold at this field is a result of the 
close proximity of a vortex near each funnel tip to the vortex in the
center of the channel combined with the asymmetry of the vortex compression
along the walls; the extra force experienced by the channel
vortex from the less compressed wall of the funnel causes it
to depin more readily.
At $B/B_{\phi} = 10$, shown in Fig.~\ref{fig:3}(c), 
two vortices occupy the center of the channel and the remaining eight
vortices line the upper and lower funnel walls in a symmetric configuration.
In Fig.~\ref{fig:3}(d) we plot the configurations at $B/B_{\phi} = 11.0$, 
where four vortices line each funnel wall and a triangular vortex structure
forms near the wide end of the funnel.
At $B/B_{\phi} = 12$, Fig.~\ref{fig:3}(e) 
shows that an asymmetric configuration of five vortices on one funnel wall and
four vortices on the other forms along with a skewed triangle of
vortices in the open portion of the channel.
At $B/B_{\phi} = 13$, the symmetric configuration illustrated
in Fig.~\ref{fig:3}(f) forms
with five vortices along each funnel wall and a triangle of vortices in the
center of the channel.

\begin{figure}
\includegraphics[width=3.5in]{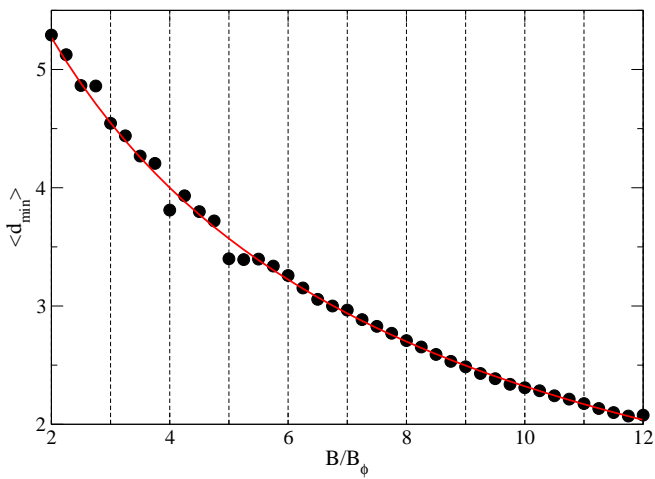}
\caption{
The average distance to the closest neighbor $\langle d_{\rm min}\rangle$
at $F_D=F_c$ vs $B/B_{\phi}$.  The line indicates a fit to $1/B$.
}
\label{fig:dmin}
\end{figure}

\subsection{Jamming and Bottleneck Effects Due to Vortex-Vortex Interactions} 

Figure~\ref{fig:1} shows that the depinning force 
tends to increase overall with increasing $B$. 
This is the result of a crowding effect that makes it more difficult for the
vortex structure to distort in order to permit individual vortices to pass
through the bottleneck.
We can show that the crowding is geometrically induced by measuring
the average closest neighbor distance $\langle d_{\rm min}\rangle$ for the
vortex configuration at the depinning force $F_D=F_c$.  To determine
$\langle d_{\rm min}\rangle$, we solve the all-nearest-neighbors problem
with a simple algorithm.  The distance from each vortex $i$ to its
closest neighbor at ${\bf R}_{nn}^{i}$ is
$d^{i}_{\rm min}=|{\bf R}_{i}-{\bf R}^{i}_{nn}|$.  Then, 
$\langle d_{\rm min}\rangle = N_v^{-1}\sum_i^{N_v}d^{i}_{\rm min}$.
The value of $\langle d_{\rm min}\rangle$ at the depinning threshold has
a convincing $1/B$ form, as shown in Fig.~\ref{fig:dmin}.  As expected in
a system where the pinning originates from vortex-vortex interaction
forces, at a given field $B$ a vortex can depin when the driving force pushes 
it closer to its neighboring vortex than the average spacing between 
vortices.  As $B$ increases, the average vortex spacing should drop as
$1/B$, consistent with the behavior of $\langle d_{\rm min}\rangle$.
The critical force curve shown in Fig.~\ref{fig:1} is very far from being
a smooth function of $B$, unlike $\langle d_{\rm min}\rangle$, and this
simply indicates that the particular geometric arrangements of the vortices
at different fields may make it easier or more difficult for two vortices
to approach each other closely enough to depin.

For the most commonly studied types of pinning, such as random pinning or
arrays of individual pinning sites, when the number of vortices exceeds the
number of pins, the depinning force tends to decrease
with increasing $B$ due to the relative increase in the strength of the
vortex-vortex interactions compared to the pinning energy.
In this case, as the vortex  lattice
becomes stiffer, some vortices are forced to shift out of the pinning
sites and occupy interstitial regions. 
In other words, a stiff vortex lattice
cannot adjust to the pinning site configuration 
as well as a soft vortex lattice can. 
One proposed mechanism for the peak effect 
observed in superconductors with random pinning 
is a softening of the vortex lattice due either to thermal fluctuations or
to changes in $\lambda$ which increase the effectiveness of the pinning.
In the funnel geometry we consider here, depinning does not require the
vortices to overcome the pinning strength of individual pinning sites.
Instead, the vortices depin once they are able to overcome the vortex-vortex
interactions blocking the passage of individual vortices through the
funnel tips.

A system that becomes less mobile due to 
increased interactions among the particles
is said to be jammed.\cite{Liu,Silbert,Drocco,Head} 
Studies on systems with short range 
interactions where a single probe particle is pushed through a collection
of other particles in the absence of pinning have shown
that the probe particle motion at a finite and 
constant driving force $F_d$ becomes
slower for increasing particle density, 
and at a critical density the probe particle becomes stuck or
jammed.\cite{Drocco,Colloid,Gazuz} 
In this jammed state, a critical driving force $F_{c}$ 
must be applied to unjam the probe particle, and this critical 
force monotonically increases with increasing 
density.\cite{Gazuz}  The critical force 
is analogous to the depinning force 
needed to move the vortices through the funnels in our system, which also
shifts to higher values as the vortex density increases.
There are many similarities between Fig.~\ref{fig:1} and
the behavior of jamming in colloidal
systems. 
At very low densities, the critical depinning force drops to zero
since the vortices are so far apart that they interact only
extremely weakly. 
Previous studies of probe particles in systems with longer range
particle-particle interactions found that a finite depinning force, similar
to a jamming effect, exists even at low densities and increases monotonically
with particle density.\cite{Hastings}
A key ingredient for jamming in our system is that a portion of the vortices
must remain immobile.
The vortices lining the walls of the funnel are held in place due to
the strength of the repulsive vortex-vortex interactions; when one of these
vortices moves toward the tip of the funnel, it experiences a barrier due
to the compression of the vortices at the funnel tip, providing a finite
depinning force.  Vortices away from the funnel walls are also pinned by means
of this compressive repulsion.
If the funnel walls did not converge, but instead remained a fixed distance
apart, the depinning force would be absent.
In experiments and simulations on straight channels 
where there are vortices both inside and outside the channels,
the vortices outside of the channel are strongly pinned
and create a periodic potential modulation 
for the vortices within 
the channel which allow the channel vortices to be  pinned.\cite{Kes}
For the funnel geometry we consider,
additional immobile vortices outside of the channel 
are not needed to create an effective periodic 
pinning potential. 
In ratchet channel geometries \cite{P}
there are strongly pinned 
vortices outside of the channels;
however, we do not expect the presence of such vortices to qualitatively
affect the results we report here.
For colloids moving through an asymmetric optical trap array, 
it would be possible to remove all colloids 
outside of the channel so that particles are present only inside the
ratchet channel.

\begin{figure}
\includegraphics[width=3.5in]{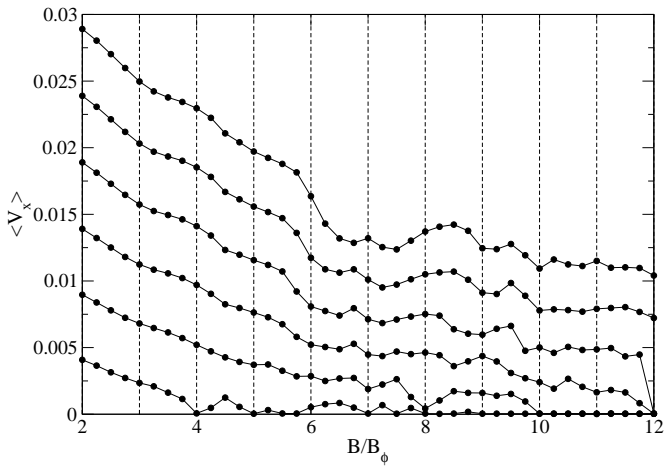}
\caption{ 
The average velocity $\langle V_x\rangle$ 
vs $B/B_{\phi}$ at
$F_{D} = 0.03$, 0.025, 0.02, 0.015, 0.01, and $0.005$ (from top to bottom). 
}
\label{fig:4}
\end{figure}

To quantify the greater difficulty with which vortices flow at higher
vortex densities,
we examine the average normalized vortex 
velocity $\langle V_x\rangle$ versus 
$B/B_{\phi}$ for different values of 
the driving force $F_{D}$.
In Fig.~\ref{fig:1} the initial depinning force $F_{c}$ was
determined by the initial onset of vortex flow;
however, the normalized velocity can be taken at any value of $F_{D}$ and in  
Fig.~\ref{fig:4} we plot $\langle V_x\rangle$ 
versus $B/B_{\phi}$ for 
$F_{D} = 0.03$, 0.025, 0.02, 0.015, 0.01, and 0.005.
In all cases, $\langle V_x\rangle$ decreases with
increasing $B/B_{\phi}$ expect for some small oscillations
caused by commensuration effects. 
At $F_D=0.005$, $\langle V_x\rangle=0$ for those fields at which $F_c>F_D$.
The reason that the overall flow velocity 
decreases for increasing $B$ is that the flow patterns organize in such a way
that only one vortex at a time is able to pass through the funnel tip, as will
be shown later.
This is a result of the very large energy cost that would be associated
with the passage of two or more vortices through the tip simultaneously.
The single passage constraint causes the
flow rate of vortices through the system to be roughly constant for
fixed $F_D$, such that for a number of moving vortices $N_m$, we find
$N_m \propto \langle V_x\rangle B$.
For fixed $N_m$, we obtain $\langle V_x\rangle \propto 1/B$,
which is approximately the behavior shown in Fig.~\ref{fig:4}. 
The results in Figs.~\ref{fig:1} and \ref{fig:4} 
show that jamming phenomena can be realized in
superconducting systems or other 
systems of particles with intermediate to long range interactions. 
Further, since stronger pinning is often a desirable feature for
many applications of superconductors, 
some of the concepts from studies of jammed systems 
could be employed to increase the effective pinning strength in a device.

\begin{figure}
\includegraphics[width=3.5in]{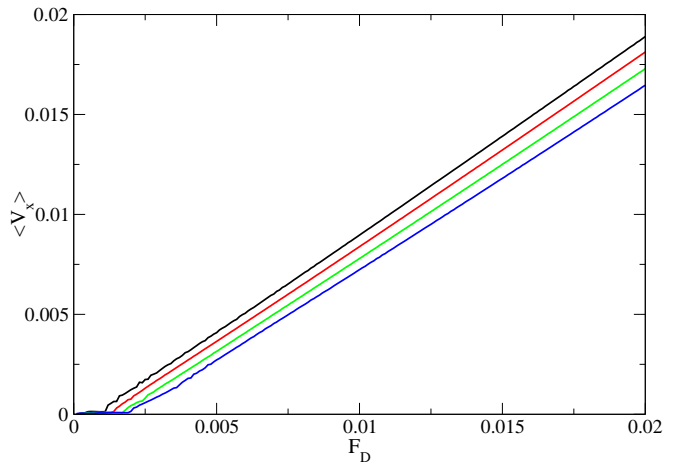}
\caption{ 
The vortex velocity $\langle V_x\rangle$ vs $F_{D}$ for 
$B/B_{\phi} = 2.0$, 2.25, 2.5, and $2.75$ (from top to bottom). 
Here the average velocity 
decreases with increasing $B$. 
}
\label{fig:5}
\end{figure}

\section{Dynamic Phases up to $B/B_{\phi} = 8.0$}

We next examine the different vortex dynamical phases that can arise in the 
funnel geometry.
In Fig.~\ref{fig:5} we plot
$\langle V_x\rangle$ versus $F_{D}$ 
for $B/B_{\phi} = 2.0$, 2.25, 2.5, and $2.75$. 
The depinning force increases with increasing $B/B_\phi$ over this
field range, as was seen in Fig.~\ref{fig:1},
and above depinning $\langle V_x\rangle$
for a particular value of $F_{D}$ decreases 
as more vortices are added to the system,
as was shown in Fig.~\ref{fig:4}.

\begin{figure}
\includegraphics[width=3.5in]{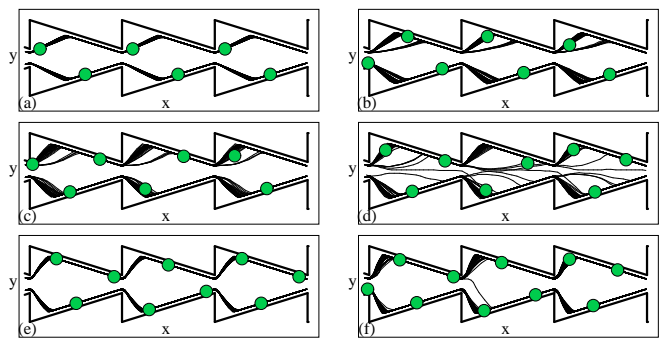}
\caption{ 
The vortex positions (dots), funnel geometry (heavy lines), 
and vortex trajectories (light lines) in a small portion of the sample
at $F_{D} = 0.01$ for $B/B_{\phi} =$  (a) 2.0, (b) $2.25$, (c) $2.5$, 
(d) $2.75$, (e) $3.0$, and (f) $3.25$. 
Here the flow at the commensurate fields
is highly ordered.  
}
\label{fig:6}
\end{figure}

In Fig.~\ref{fig:6} we plot the vortex trajectories at fixed $F_D=0.01$ for
$B/B_{\phi} = 2.0$, 2.25, 2.5, 2.75, 3.0,
and $3.25$. 
The flow is ordered at $B/B_{\phi}  = 2.0$ and $B/B_{\phi}=3.0$ 
in Fig.~\ref{fig:6}(a) and Fig.~\ref{fig:6}(e). 
The vortices move in 
fixed trajectories and continuously maintain the same neighbors, 
indicating that all of the vortices are moving.
For $B/B_{\phi} = 2.25$, 2.5, and $2.75$ 
in Fig.~\ref{fig:6}(b,c,d), the trajectories become increasingly disordered and 
the vortices no longer keep the same neighbors 
over time, indicative of plastic flow. 
The disordered nature of the flow permits the
vortices to explore larger regions of phase space, 
including regimes in which some of the vortices move much more slowly than
others or become temporarily pinned.

\begin{figure}
\includegraphics[width=3.5in]{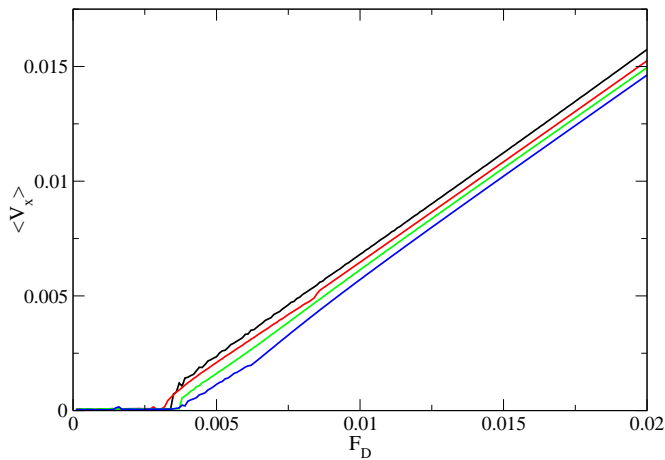}
\caption{ 
$\langle V_x\rangle$ 
vs $F_{D}$ for $B/B_{\phi} = 3.0$, 3.25, 3.5, and $3.75$ (from top right
to bottom right), 
showing the
same trends as in Fig.~5.   
}
\label{fig:7}
\end{figure}

Fig.~\ref{fig:6}(f) indicates that at $B/B_{\phi}=3.25$, the  
vortex trajectories are 
more disordered than at $B/B_{\phi} = 3.0$. 
In  Fig.~\ref{fig:7} we plot $\langle V_x\rangle$ vs $F_{D}$  
for $B/B_{\phi} = 3.0$, 3.25, 3.5, and $3.75$,
which show the same trend observed at and above the second matching field. 
The value of $\langle V_x\rangle$ at a fixed $F_{D}$ above the depinning
threshold decreases with
increasing $B/B_{\phi}$ and the 
vortex flow becomes partially disordered at the incommensurate fields.         

\begin{figure}
\includegraphics[width=3.5in]{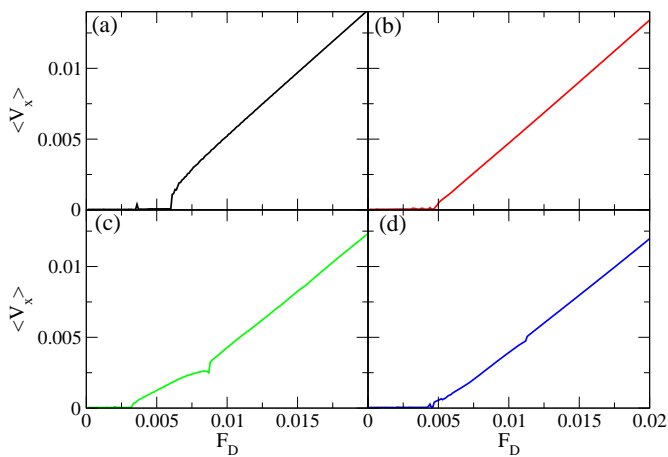}
\caption{ 
$\langle V_x\rangle$ 
vs $F_{D}$ for 
(a) $B/B_{\phi} = 4.0$. 
(b) $B/B_\phi=4.25$. 
(c) $B/B_\phi=4.5$ showing a two step
depinning process with a cusp near 
$F_{D} = 0.008$ where $\langle V_x\rangle$ decreases with
increasing $F_{D}$. 
(d) $B/B_\phi=4.75$ showing some small steplike features.   
}
\label{fig:8}
\end{figure}

\begin{figure}
\includegraphics[width=3.5in]{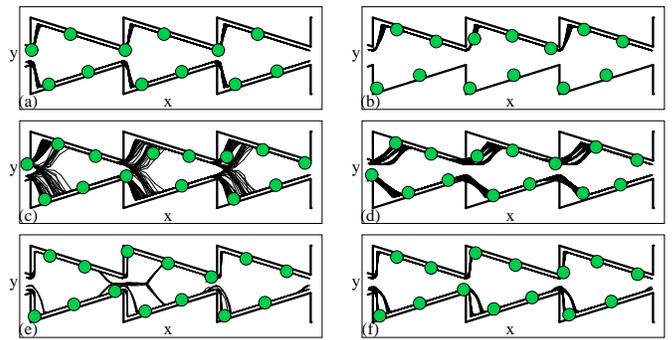}
\caption{ 
Vortex portions (dots), funnel geometry (heavy lines), and 
vortex trajectories (light lines) in a small portion of the sample. 
(a) $B/B_{\phi} = 4.0$ at $F_D=0.01$ where an ordered flow occurs. 
(b) The initial ordered flow at $B/B_{\phi} = 4.5$ 
and $F_D=0.008$, below the cusp in 
$\langle V_x\rangle$ in Fig.~\ref{fig:8}(c), where only the vortices
along the top wall are moving.
(c) The disordered flow for $B/B_\phi=4.5$ at $F_D=0.017$, 
above the cusp in 
$\langle V_x\rangle$ in Fig.~\ref{fig:8}(c), 
where all of the vortices are moving. 
(d) The high drive phase at $F_{D} = 0.0275$ for $B/B_{\phi} = 4.5$.
(e) The initial disordered flow at 
$F_D=0.008$ for $B/B_{\phi} = 4.75$. 
(f) The flow at $F_D=0.01$ for $B/B_\phi=4.75$, below the step 
in $\langle V_x\rangle$ which occurs
near $F_{D} = 0.012$ in Fig.~\ref{fig:8}(d). 
For drives above the step in $\langle V_x\rangle$ at this field,
the vortex flow resembles the flow shown in panel (a).    
}
\label{fig:9}
\end{figure}

For $B/B_{\phi} > 4.0$, the transport becomes more complicated and transitions
between distinct dynamical phases begin to occur.
The moving phases for $B/B_{\phi} > 4.0$ are 
generally characterized as plastic since a portion of the vortices
can remain immobile.  We find both
ordered plastic motion, where the trajectories of the
mobile vortices follow a fixed path, 
and disordered plastic flow phases 
where the vortex motion is more chaotic and the
vortices follow many different paths.
In Fig.~\ref{fig:8}(a-d) we plot
$\langle V_x\rangle$ 
versus $F_{D}$ for $B/B_{\phi} = 4.0$, 4.25, 4.5, and 4.75.  At 
$B/B_{\phi} = 4.0$, Fig.~\ref{fig:8}(a) indicates 
that there is a single well defined depinning transition 
where the vortices flow
elastically, as illustrated in Fig.~\ref{fig:9}(a). 
For $B/B_{\phi} = 4.25$, the value of $F_c$ is depressed as seen in
Fig.~\ref{fig:8}(b). Here there are no sharp jumps in the transport curve, 
and the vortex motion resembles that   
observed at $B/B_{\phi} = 4.0$ but with some additional
fluctuations in the vortex trajectories. 

At $B/B_{\phi} = 4.5$, a two step depinning process occurs as
shown in Fig.~\ref{fig:8}(c).
Initially, only one 
ordered channel of moving vortices forms along one of the 
funnel walls
while the remaining vortices are immobile.  This is
illustrated in Fig.~\ref{fig:9}(b) for a simulation in which the vortices along the
upper funnel wall depinned first; depending upon the initial random 
fluctuations, it is also possible for the vortices along the lower wall to
depin first.
The remaining vortices depin near $F_D=0.008$, where a cusp feature
appears in $\langle V_x\rangle$. 
At the cusp, the average vortex velocity decreases with
increasing $F_{D}$, 
creating
a region of negative differential conductivity where 
$d\langle V_x\rangle/dF_{D} < 0.0$. 
The negative differential conductivity at 
$B/B_{\phi} = 4.5$ is not as pronounced as that observed in simulations
and experiments with square pinning arrays; 
\cite{Reichhardt,Misko} however, we find that the type 
of negative differential conductivity illustrated in Fig.~\ref{fig:8}(c) 
is a common feature at a number of the 
higher order incommensurate fillings we have examined
in the funnel geometry. 
In contrast, for the square pinning array the 
negative differential conductivity occurs only near the first matching 
field.  This indicates that it may actually be easier to observe
negative differential conductivity in a funnel geometry than in a square
pinning array.
Above the cusp at $F_D=0.008$, Fig.~\ref{fig:8}(c) shows that $\langle V_x\rangle$ 
smoothly increases with increasing $F_{D}$. 

In Fig.~\ref{fig:9}(b) we show that the vortex flow below the cusp 
in $\langle V_x\rangle$ at 
$B/B_{\phi} = 4.5$ and $F_D=0.008$ is ordered with only a single channel of
vortex flow. 
Figure \ref{fig:9}(c) indicates that 
for the same field at $F_D=0.017$, above the cusp 
in $\langle V_x\rangle$,
the vortices flow in disordered paths, each of which predominantly runs
along either the upper or lower funnel wall.  The transition from
only a single flowing channel to effectively two flowing channels of
vortices at the second depinning transition associated with the cusp 
in $\langle V_x\rangle$
causes a drop in the mobility of the vortices due to the competition
between the two channels for 
passing a vortex through the
bottleneck of the funnel.  Only a single vortex can fit through the
bottleneck at a time, but vortices in the upper and lower flowing channels
do not arrive at the bottleneck at synchronized times due to the unequal
distribution of vortices between the two effective channels
which causes the two channels to flow at different average speeds.  
As a result,
a vortex in one channel may reach the bottleneck too
soon while a vortex from the other channel is still moving through the
bottleneck, forcing the vortices in the first channel to move more
slowly until the vortex in the second channel has exited the bottleneck
and freed it for passage of a vortex in the first channel.
As $F_D$ increases, the difference in flow speed for the two unequally
populated channels of moving vortices decreases until it is small enough
that the flow of the two channels becomes synchronized on average and
a much more orderly passage of vortices through the bottleneck
occurs, alternating between the two channels.
In this case, the disordered trajectories become smoother, as
shown for $F_{D} = 0.0275$ in Fig.~\ref{fig:9}(d). 

For $B/B_{\phi} = 4.75$, Fig.~\ref{fig:8}(d) shows that
there are several small steps in the 
$\langle V_x\rangle$ versus $F_{D}$ curve which correspond
to changes in the flow. 
In Fig.~\ref{fig:9}(e), the initial flow for this field
at $F_D=0.008$ is partially disordered. 
By $F_D=0.01$, shown in Fig.~\ref{fig:9}(f), a transition
to a more ordered flow state 
has occurred where there are two possible paths for vortices to follow
in the lower channel. 
At higher drives, one of these two lower paths closes,  
corresponding to the jump in $\langle V_x\rangle$ near $F_{D} = 0.012$, 
and the flow at higher drives resembles that shown in Fig.~\ref{fig:9}(a).   

\begin{figure}
\includegraphics[width=3.5in]{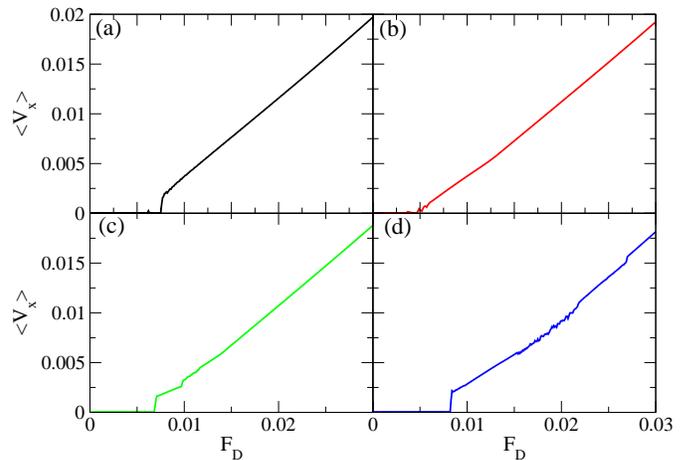}
\caption{ 
$\langle V_x\rangle$ vs $F_{D}$ for 
(a) $B/B_{\phi} = 5.0$. 
(b) $B/B_\phi=5.25$. 
(c) $B/B_\phi=5.5$. 
(d) $B/B_\phi=5.75$.
}
\label{fig:10}
\end{figure}

\begin{figure}
\includegraphics[width=3.5in]{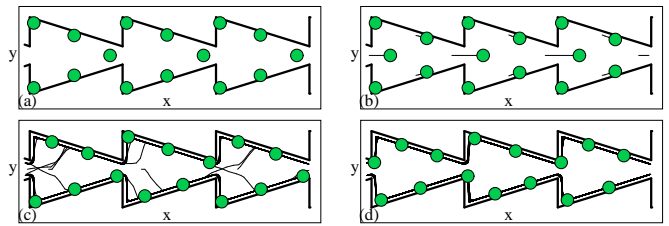}
\caption{ 
The structural transition that occurs in the single stage depinning
transition at $B/B_{\phi}=5.0$ shown in a small portion of the sample. 
Dots: vortex positions; heavy lines: funnel geometry; light lines: vortex
trajectories.
(a) Right before depinning, at $F_D=0.006$, there are four
vortices lining the walls and one vortex located 
in the center of the channel near the funnel tip.
(b) The first rearrangement just below $F_D=0.0075$ 
occurs when the center vortices move through the aperture in the positive
$x$ direction and pass into the adjacent funnels.     
(c) At depinning, which occurs at $F_D=0.0075$, 
these center vortices move against the 
funnel walls in an alternating pattern associated with a transient flow. 
(d) Above depinning
at $F_D=0.012$, the vortices flow strictly along the funnel walls.   
}
\label{fig:11}
\end{figure}

At $B/B_{\phi} = 5.0$ there is a single depinning transition, as illustrated
in the plot of $\langle V_x\rangle$ versus $F_D$ 
in Fig.~\ref{fig:10}(a). 
Right at the
depinning transition, 
the vortices undergo a structural transition which is highlighted in 
Fig.~\ref{fig:11}. The vortex configuration for drives well below depinning
consists of four vortices arranged symmetrically along the funnel walls 
with a fifth vortex in the center of the channel near the
funnel tip, as shown in Fig.~\ref{fig:11}(a). 
At the onset of depinning, illustrated in 
Fig.~\ref{fig:11}(b) just below $F_D=0.0075$,
the vortices at the funnel tips move in the  
positive $x$ direction and shift into the adjacent funnel. 
At depinning, which occurs at $F_D=0.0075$, Fig.~\ref{fig:11}(c) 
indicates the
rearrangement that occurs when the center vortices move against one of
the funnel walls in an alternating pattern.
Above depinning, as shown in Fig.~\ref{fig:11}(d) for $F_D=0.012$, the vortices flow
strictly along the funnel walls in two channels that never mix.

\begin{figure}
\includegraphics[width=3.5in]{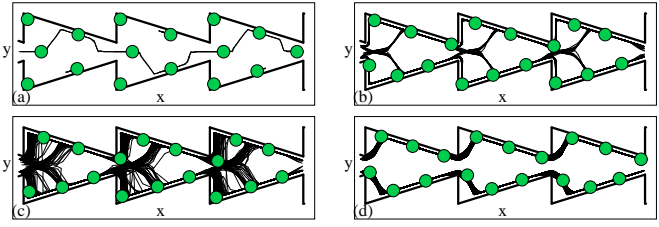}
\caption{ 
The vortex positions (dots), funnel geometry (heavy lines), and vortex
trajectories (light lines) in a small portion of the sample.
(a) The initial flow at $F_D=0.006$ for $B/B_{\phi} = 5.25$. 
(b) The initial depinning at $F_D=0.012$ for $B/B_{\phi} = 5.75$. 
(c) The trajectories in the fluctuating phase for 
$F_D=0.019$ at $B/B_{\phi} = 5.75$. 
(d) The high drive ordered phase at $F_D=0.032$ for $B/B_{\phi} = 5.75$.   
}
\label{fig:12}
\end{figure}

At $B/B_{\phi} = 5.25$, 5.5, and 5.75, shown in Fig.~\ref{fig:10}(b), 
\ref{fig:10}(c), and \ref{fig:10}(d), 
respectively, there can be multiple
dynamical transitions between ordered and disordered flow phases. 
At $B/B_{\phi} = 5.25$, the flow at large $F_{D}$ is very similar to that shown
for $B/B_{\phi} = 5.0$ in Fig.~\ref{fig:11}(d); 
however, the flow initiates at a much
lower drive and takes the form of an ordered
winding channel in which about 40\% of the vortices
are moving, 
as shown in Fig.~\ref{fig:12}(a). 
As the vortices in the winding channel move through the system, the vortices
pinned along the funnel walls undergo an oscillatory motion which is most
easily seen in Fig.~\ref{fig:12}(a) for the pinned vortex near the center of each
long funnel wall. 
This oscillation is a response to the passage of an individual vortex
through the flowing channel; the pinned vortex shifts in the positive
$x$ direction as the
moving vortex approaches, and shifts back again after the moving vortex
has passed.
The vortices at the corners of the funnel undergo little to no shift. 
In Fig.~\ref{fig:10}(d) 
at $B/B_{\phi} = 5.75$ the transitions between different dynamical phases 
are associated with changes in the fluctuations of the velocity signal.
At this filling, the sharp depinning transition takes the system into
the ordered braiding flow phase illustrated in Fig.~\ref{fig:12}(b), a state with
low levels of velocity fluctuations. 
A transition to a highly fluctuating phase
occurs near $F_{D} = 0.15$ and corresponds 
to the onset of the disordered flow phase illustrated in Fig.~\ref{fig:12}(c).  
For higher drives, a transition to an ordered phase occurs near 
$F_{D}=0.0225$ when the trajectories become partially ordered. 
A velocity jump near $F_D=0.025$ marks the transition to the completely
ordered flow phase shown in
Fig.~\ref{fig:12}(d). 

\begin{figure}
\includegraphics[width=3.5in]{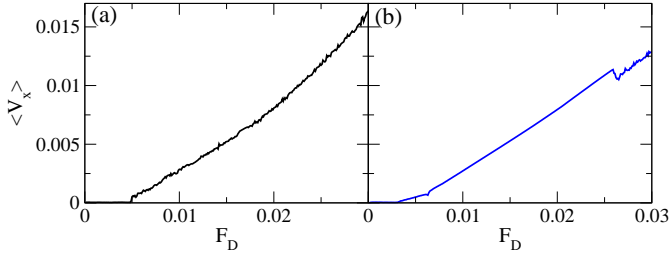}
\caption{ 
$\langle V_x\rangle$ vs $F_{D}$ for (a) $B/B_{\phi} = 6.0$. 
(b) $B/B_{\phi} = 6.75$.     
}
\label{fig:13}
\end{figure}

\begin{figure}
\includegraphics[width=3.5in]{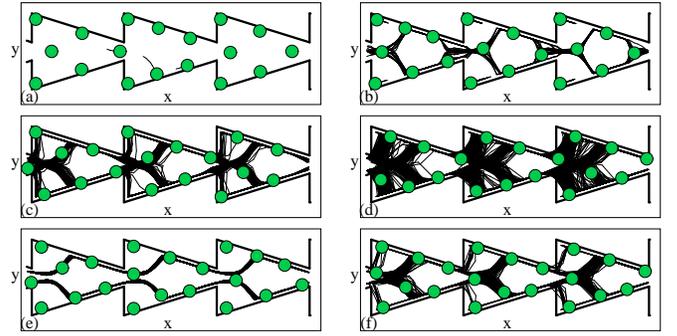}
\caption{ 
The vortex positions (dots), funnel geometry (heavy lines), and vortex 
trajectories (light lines) in a small portion of the sample.
(a) The initial motion at $F_D=0.006$ for $B/B_{\phi} = 6.0$.
(b) The flow above depinning at $F_D=0.008$ for $B/B_{\phi} = 6.0$.
(c) The flow at $F_D=0.019$ for $B/B_{\phi} = 6.0$. 
(d) The flow at $F_D=0.028$ for $B/B_{\phi} = 6.0$.
(e) The ordered vortex flow at $F_D=0.025$ 
for $B/B_\phi=6.75$, below the cusp in $\langle V_x\rangle$ vs 
$F_{D}$ seen in Fig.~\ref{fig:13}(b).
(f) The disordered flow at $F_D=0.028$ for $B/B_\phi=6.75$, just after the cusp 
in $\langle V_x\rangle$.
}
\label{fig:14}
\end{figure}

At $B/B_{\phi}= 6.0$, where a commensuration peak in $F_c$ was missing
in Fig.~\ref{fig:1},
the velocity-force curve shows a single 
depinning transition into a random flow phase, as shown in Fig.~\ref{fig:13}(a). 
No sharp transitions between different phases appear; however, the general
form of the vortex flow changes gradually as $F_D$ increases.
Figure~\ref{fig:14}(a) shows an initial shift in vortex positions occurring at a drive
just below the depinning transition.
The vortex at the tip of the leftmost funnel 
has pushed the vortex at the base of the adjacent funnel into a position
along the lower wall in a pattern that repeats every two funnel plaquettes, 
indicating that the vortices in the center of the channel
were not well pinned, resulting in 
lack of a peak in $F_{c}$ at this field. 
Just above depinning, shown at $F_D=0.008$ in Fig.~\ref{fig:14}(b), 
a disordered flow occurs in which the two vortices at the corners
of each funnel do not participate.
Although the flow is disordered,
there are clearly defined regions inside the funnels which the vortices
completely avoid, as shown by the lack of trajectories passing through
large areas of the funnels. 
As $F_{D}$ is further increased, 
the trajectories become increasingly disordered and 
the vortices in the corners of the funnels begin to 
take part in the motion,
as shown in Figs.~\ref{fig:14}(c,d).      

More clearly defined transitions between ordered and disordered flow
states occur at $B/B_{\phi} = 6.25$, 6.5, and 6.75, as illustrated 
in Fig.~\ref{fig:13}(b) for $B/B_\phi=6.75$.
The transitions are characterized by cusp structures 
in $\langle V_x\rangle$
associated with negative differential conductivity.
Fig.~\ref{fig:14}(e) shows the initial ordered flow 
at $F_D=0.028$ for $B/B_\phi=6.75$, below a pronounced cusp
in $\langle V_x\rangle$. Here the two vortices in the
corners of each funnel remain immobile 
so that the flow is actually plastic. 
These vortices become mobile at 
the transition to the disordered phase, illustrated at $F_D=0.028$ in 
Fig.~\ref{fig:14}(e). 
A similar set of dynamics appears at $B/B_{\phi} = 6.25$ and 
$B/B_\phi=6.5$. 
For driving forces higher than those we have examined,
it is possible that further dynamical 
transitions could occur at which different ordered phases arise.  

\begin{figure}
\includegraphics[width=3.5in]{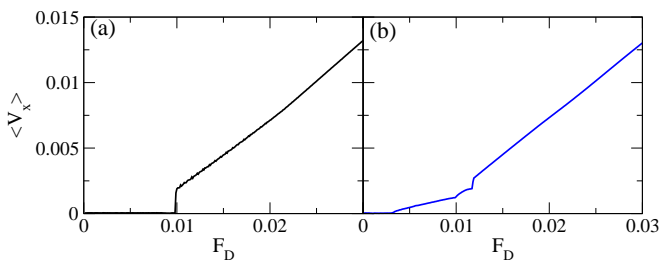}
\caption{ 
$\langle V_x\rangle$ vs $F_{D}$ for (a) $B/B_{\phi} = 7.0$
and (b) $B/B_{\phi} = 7.75$, where three distinct moving phases
appear.    
}
\label{fig:15}
\end{figure}

\begin{figure}
\includegraphics[width=3.5in]{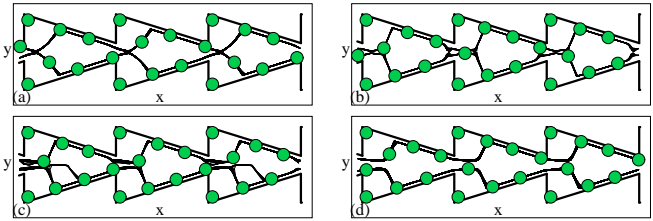}
\caption{ 
The vortex positions (dots), funnel geometry (heavy lines), and
vortex trajectories (light lines) in a small portion of the sample.  
(a) The vortex flow state at $F_D=0.014$ for $B/B_{\phi} = 7.0$.
(b) The braided flow phase at $F_D=0.008$ for $B/B_{\phi} = 7.75$.
(c) The second flow phase at $F_D=0.011$ for $B/B_{\phi} = 7.75$.
(d) The third flow phase at $F_D=0.021$ for $B/B_{\phi} = 7.75$.   
}
\label{fig:16}
\end{figure}

Depinning at $B/B_\phi=7.0$ occurs in a single step, as illustrated in 
Fig.~\ref{fig:15}(a) where we plot $\langle V_x\rangle$ versus $F_D$.  At this
field, Fig.~\ref{fig:1} indicates that there is a peak in $F_c$.
Above depinning, the vortices flow in the ordered pattern shown
in Fig.~\ref{fig:16}(a), where two vortices remain pinned in the corners
of each funnel. 
For drives $F_D>0.3$ higher than those we consider here,
a transition to a disordered flow state may occur 
once the drive is large enough to cause
the two immobile vortices to depin.
At $B/B_{\phi} = 7.25$, 7.5, and $7.75$, multiple
steps occur in the velocity-force curves, 
as illustrated in Fig.~\ref{fig:15}(b) for $B/B_{\phi} = 7.75$. 
Each step is associated with a distinct type of ordered flow.
At low drives, we find the braided flow shown in Fig.~\ref{fig:16}(b) 
at $F_D=0.008$.
Above the first low step in $\langle V_x\rangle$ near $F_D=0.01$, 
an alternating braided flow 
occurs in which every other
plaquette contains two possible flow paths
along the lower funnel wall, as illustrated
in Fig.~\ref{fig:16}(c) at $F_D=0.011$.
Above the second larger step
in $\langle V_x\rangle$ near $F_D=0.012$, 
the completely ordered flow phase shown
in Fig.~\ref{fig:16}(d) at $F_D=0.021$ occurs.

\begin{figure}
\includegraphics[width=3.5in]{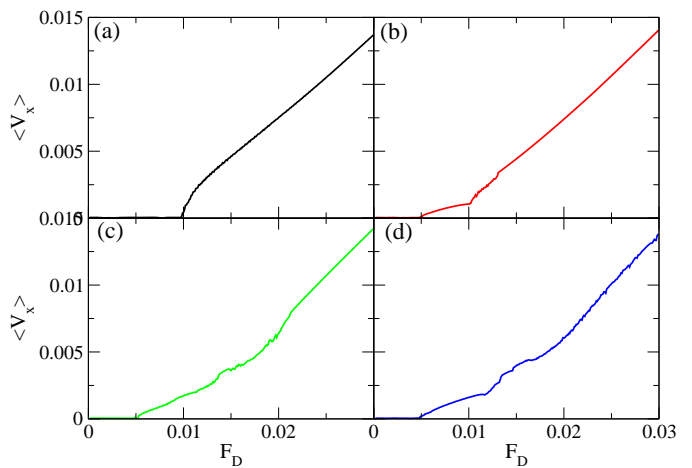}
\caption{ 
$\langle V_x\rangle$ vs $F_{D}$ for 
(a) $B/B_{\phi} = 8.0$, (b) $B/B_\phi=8.25$, (c) $B/B_\phi=8.5$,
and (d) $B/B_\phi=8.75$.
}
\label{fig:17}
\end{figure}

\begin{figure}
\includegraphics[width=3.5in]{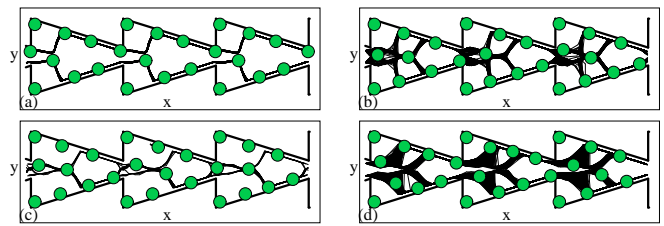}
\caption{ 
The vortex positions (dots), funnel geometry (heavy lines), and
vortex trajectories (light lines) in a small portion of the sample. 
(a) The ordered flow phase at $F_D=0.014$ for $B/B_{\phi} = 8.0$.
(b) The disordered flow at $F_D=0.017$ for $B/B_{\phi} = 8.5$.
(c) The initial partially ordered flow phase at 
$F_D=0.008$ for $B/B_{\phi} = 8.75$.
(d) The disordered flow phase with two immobile vortices in the corners
of each funnel at $F_D=0.025$ for $B/B_{\phi} = 8.75$.   
}
\label{fig:18}
\end{figure}

\section{Dynamics at Higher fields $B/B_\phi\geq 8$}  

For higher fields $B/B_\phi\geq 8.0$ we observe the same trends 
found for fields just below $B/B_{\phi} = 8.0$. 
The integer matching fields typically display
a single ordered flow phase, 
while at the incommensurate fields, multiple flow phases occur with
ordered-ordered or ordered-disordered flow transitions.
Another trend is that as the field increases, the number of immobile vortices
appearing in the ordered flow phases increases.
In Fig.~\ref{fig:17} we plot $\langle V_x\rangle$ versus $F_{D}$
for $B/B_{\phi} = 8.0$, 8.25, 8.5, and 8.75, where the trend outlined above
can be seen. 
There is a single depinning step in Fig.~\ref{fig:17}(a) at $B/B_{\phi} = 8.0$ 
into the ordered flow phase shown in Fig.~\ref{fig:18}(a) at $F_D=0.014$
where two vortices remain pinned at the corners of each funnel. 
Figure~\ref{fig:18}(b) shows the disordered flow at the higher drive $F_D=0.017$
for $B/B_{\phi} = 8.5$.  At this field, 
there is a transition near $F_D=0.025$ to an ordered phase 
with vortex trajectories that are very similar 
to those shown in Fig.~\ref{fig:18}(a) for $B/B_{\phi} = 8.0$.
The initial flow at $B/B_{\phi} = 8.75$ 
is partially ordered with four immobile vortices 
in each funnel plaquette, as shown at $F_D=0.008$ 
in Fig.~\ref{fig:18}(c).
At higher drives for this field, the flow becomes disordered 
but only two of the immobile vortices begin to move, leaving two vortices
immobile in the corners of each funnel, 
as shown in Fig.~\ref{fig:18}(d) for $F_D=0.025$. 

\begin{figure}
\includegraphics[width=3.5in]{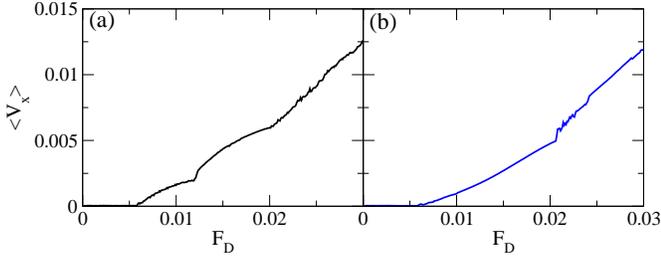}
\caption{ 
$\langle V_x\rangle$ vs $F_{D}$ for (a) $B/B_{\phi} = 9.0$ and  
(b) $B/B_{\phi} = 9.75$. 
}
\label{fig:19}
\end{figure}

\begin{figure}
\includegraphics[width=3.5in]{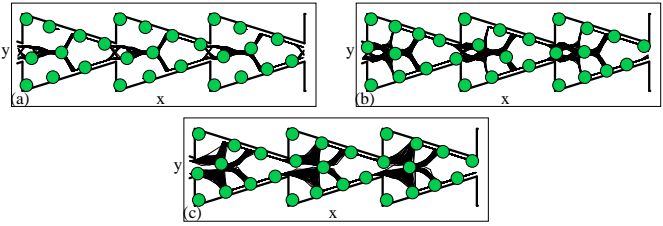}
\caption{
The vortex positions (dots), funnel geometry (heavy lines), and vortex
trajectories (light lines) 
in a small portion of the sample at $B/B_\phi=9.0$ showing braided flows.
(a) First flow regime shown at $F_D=0.01$.
(b) Second flow regime shown at $F_D=0.014$.
(c) Third flow regime shown at $F_D=0.023$.
}
\label{fig:20}
\end{figure}

\begin{figure}
\includegraphics[width=3.5in]{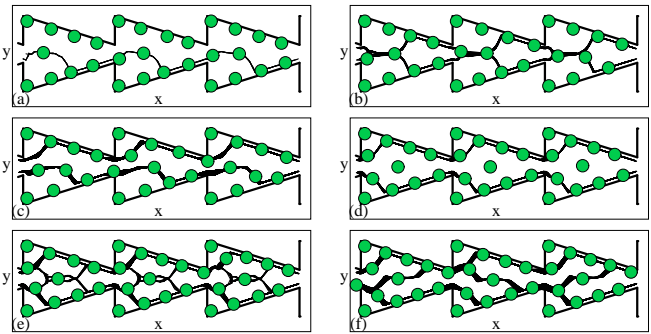}
\caption{ 
The vortex positions (dots), funnel geometry (heavy lines), and  
vortex trajectories (light lines) in a small portion of the sample.
(a) The initial motion at $F_D=0.008$ for $B/B_{\phi} = 9.75$.
(b) The second flow phase at $F_D=0.012$ for $B/B_{\phi} = 9.75$. 
(c) The higher drive ordered flow phase at $F_D=0.028$ for 
$B/B_{\phi} = 9.75$.     
(d) The flow at $F_D=0.017$ for $B/B_{\phi} = 11.0$. 
(e) A complex partially ordered flow pattern at 
$F_D=0.017$ for $B/B_{\phi} = 11.25$.
(f) A higher drive ordered alternating flow pattern at 
$F_D=0.025$ for $B/B_{\phi} = 11.75$. 
}
\label{fig:21}
\end{figure}

We find a distinctive integer matching field at which there is not merely 
a single dynamic flow phase state.  This field, $B/B_\phi=9.0$, also fails
to produce a depinning peak as shown in Fig.~\ref{fig:1}.
We plot $\langle V_x\rangle$ versus $F_D$ for $B/B_\phi=9.0$ in 
Fig.~\ref{fig:19}(a),
which indicates that there are three distinct moving phases.
All three phases involve braided vortex flows, but the details of the
braiding and the number of vortices participating in the flow varies as
a function of drive, as shown in Fig.~\ref{fig:20} where the three different
braided flows are illustrated.
The suppression of the depinning peak at $B/B_\phi=9.0$ is related to
the suppression of the peak at $B/B_\phi=6.0$.  
At each of these fields, the system exhibits an instability
in which the number of vortices present in the center of the channel for
zero applied drive is different than the number of vortices that can be
stabilized along the center of the channel just above depinning.  At
$B/B_\phi=6$, in each funnel plaquette there are two vortices positioned along 
the center line of the channel at zero drive, but under finite drive one
of these vortices moves against the funnel wall, leaving only one vortex
in the center of the channel.  This transition can be seen by comparing
Fig. 2(e) with Fig. 15(a,b).  At this filling, there is only a small energy
difference between placing two vortices at the center of the channel and
placing only one vortex at the center of the channel with the other vortex
against the funnel wall.  This produces an instability which contributes to
the disordered flow we find at $B/B_\phi=6$.  An opposite transition occurs
for $B/B_\phi=9$.  As seen in Fig. 3(b) and Fig. 21(a), there is only one
vortex per plaquette in the center of the channel at zero applied drive, but
above depinning a second vortex moves into the center of the channel and
is stabilized there.  The existence of this change in the occupancy of the
center of the channel at depinning from one to two or two to one
is associated with the suppression of commensuration peaks in the critical
depinning force at the fields $B/B_\phi=6$ and $B/B_\phi=9$ seen in
Fig. 1.  

A series of flow transitions also occurs at
$B/B_\phi=9.75$ in Fig.~\ref{fig:19}(b), corresponding to the dynamical
flows illustrated in Fig.~\ref{fig:21}(a,b,c).
In the initial flow phase at low drives, Fig.~\ref{fig:21}(a) 
indicates that only
a portion of the vortices are moving in a path that passes through the bottom
half of each funnel, forming a single flowing channel.
At higher drives a transition to the symmetric flow state 
illustrated in Fig.~\ref{fig:21}(b) occurs. 
At still higher drives, an intermediate disordered phase appears before
the flow reorders into the state shown for $F_D=0.028$ in Fig.~\ref{fig:21}(c),
where there are two immobile vortices on the bottom half of each funnel
and one immobile vortex on the top half of the funnel, with flow
occurring through two well-defined channels.

At higher fields, similar dynamical regimes occur 
and larger numbers of vortices can become immobile along the funnel walls,
such as at $B/B_{\phi} = 10$ 
where there is a single step depinning transition into a state with
four immobile vortices in each funnel.
It is also possible for vortices to be immobilized in the center of the
channel rather than along the funnel walls, as illustrated for
$B/B_\phi=11.0$ in Fig.~\ref{fig:21}(d).
At this field, a single step depinning transition 
occurs into the state shown where a total of three vortices are immobilized:
two in the corners of the funnel and one 
right in the funnel center.
As $B/B_{\phi}$ increases, 
additional partially ordered phases occur with increasingly intricate 
flow patterns such 
as the one shown in Fig.~\ref{fig:21}(e) at $B/B_{\phi} = 11.25$. 
We also find phases with asymmetric flow where the asymmetry alternates
from one funnel to the next, such as the state illustrated
in Fig.~\ref{fig:21}(f) for $B/B_{\phi} = 11.75$. 
Here the leftmost funnel contains two flowing
channels of vortices interacting with the upper funnel wall and one 
channel interacting with the lower funnel wall, while in the
middle funnel this flow pattern is reversed.

We expect the trends described above to continue for higher matching 
fields.  Eventually, however, there may be a crossover to a state
where the vortex lattice constant is smaller than the 
width of the funnel aperture, at which point it 
would be possible to move two vortices through the funnel tip at 
the same time without much of an additional energy cost.
In this case, the average velocity might sharply increase since the
flow in the aperture would transition from one-dimensional to 
quasi-two-dimensional,
and it may be possible that further
oscillations in the average velocity would arise 
at even higher fields as larger groups of vortices could pass through
the aperture simultaneously for increasing magnetic field values.

\begin{figure}
\includegraphics[width=3.5in]{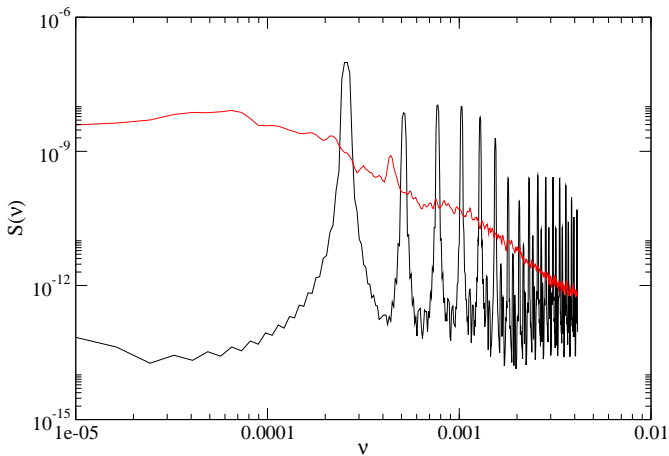}
\caption{
The power spectrum $S(\nu)$ versus frequency $\nu$ in inverse simulation
time steps.  Lower curve: The ordered phase at 
$F_D=0.02$ for $B/B_{\phi} = 5.0$, where 
a series of peaks occurs due to the periodic velocity signal 
of the vortices.  Upper curve: The disordered flow phase
at $F_D=0.02$ for $B/B_{\phi} = 6.0$, 
where a broadband noise signal arises with 
a $1/\nu^{2.5}$ form.  The same types of spectra appear for 
ordered and disordered flow phases at other fields.  
}
\label{fig:22}
\end{figure}

\section{Noise measurement and Temperature Effects}

With current experimental noise measurement techniques,
it should be possible to characterize the difference
between ordered and disordered flow phases. 
The vortex velocities we measure in our simulations are 
proportional to the experimentally measured
vortex voltage signal, which can be analyzed using the power spectrum
defined as
\begin{equation}
S(\nu) = \left|\frac{1}{\sqrt{2\pi}}\int V(t)e^{-i 2\pi\nu t} dt\right|^2  
\end{equation}
In Fig.~\ref{fig:22} we plot $S(\nu)$ for $B/B_{\phi} = 5.0$ 
at $F_D=0.02$ in a regime of ordered flow. We find narrow band noise peaks
at the characteristic frequencies of the ordered oscillatory motion of the
vortices.
Also plotted in Fig.~\ref{fig:22} is $S(\nu)$ for $B/B_{\phi} = 6.0$
at $F_D=0.02$, where the
vortex flow is disordered. 
Here the power spectrum has a broad band or $1/f^{\alpha}$ 
noise characteristic with 
$\alpha = 2.5$. 
In general, for other values of $B/B_{\phi}$, 
the ordered phases produce narrow band spectra while the
disordered phases produce broad band noise signals. 
The appearance of the narrow band noise in the ordered phase
implies that phase locking phenomena could be induced by adding an ac
drive component to the dc driving force.
When the frequency of the ac drive matches the intrinsic 
or higher harmonic frequencies of the
moving vortices in the ordered phase, 
a series of steps should appear 
in the velocity force curves.    

We next consider the effect of temperature. 
In periodic pinning arrays, the most pronounced matching field peaks
observed in experiments occur at temperatures near $T_{c}$. 
This has been attributed to the thermal suppression of the 
intrinsic pinning in the sample that competes with the periodic pinning. 
Another factor that could play a role is that the vacancies or interstitials
present in the vortex lattice away from commensurate fields can be 
relatively mobile under thermal activation, and so increasing the
temperature could depress the depinning force to a larger degree at 
incommensurate fields than at commensurate fields where vacancies or
interstitials are not present.
We find that for a temperature of $T=0.1$, which is well below the melting 
temperature of the vortex lattice,
the depinning force peaks at the commensurate fields are robust 
while the depinning force at the incommensurate fields 
is depressed. The peaks in $F_c$ that appear at 
$B/B_{\phi} = 4.5$ and $B/B_\phi=5.5$ for 
$T = 0$ are absent at $T=0.1$, indicating that for finite temperatures,
pronounced peaks are only observable at the matching fields whereas submatching
field peaks are less robust.
The general trend of increasing $F_{c}$ with
increasing $B/B_{\phi}$ also remains robust at finite temperature.  

\section{Discussion}
Although our work is focused on a superconducting vortex system, 
we expect our results to be general for
other systems of particles with repulsive interactions. 
For example, in colloidal systems where the volume density is sufficiently
low that steric contact of the colloidal particles does not occur, many of
the same results should still apply.
One aspect of the vortex system that we do not take 
into account is the possibility of multi-quanta vortex formation under
certain conditions.
In static systems, the formation of giant vortices has been observed
in cases where the giant vortices produce a higher symmetry of the
vortex configuration, such as in quasiperiodic pinning arrays.\cite{Mosh2}
It is possible that dynamical multi-quanta vortices could form in a 
funnel geometry.  For example, if two vortices are forced into close 
proximity near the funnel tip, they could merge to form a two-quanta vortex
which would then pass through the funnel tip.
The formation of such dynamical multi-quanta 
vortices could produce interesting signatures in the transport curves. 
The experimental system closest to the system studied here
is periodic asymmetric channel geometries \cite{P} 
where small oscillations in the critical current due to commensuration
effects were observed.  Since these commensuration effects were very 
weak, it is possible that they were due to edge effects.
More recent
experiments with geometries that avoid the effect of edges 
have now revealed much more prominent commensuration
effects,\cite{RatchetN} 
indicating that some of the phases we observe may be occurring 
in this system.      

\section{Summary} 
In summary, we have used numerical simulations 
to examine the vortex configurations and dynamics
in a periodic funnel array. 
The vortex configurations we 
observe are generally different from those 
found for a single isolated
triangular sample due to the coupling between vortices in
adjacent funnels.
As a function of field we find a series of depinning threshold peaks at 
the matching fields where the vortex configurations are ordered. 
In some cases, matching peaks are missing due to the fact that the
vortex configuration contains pairs of vortices that are located
close together near the funnel apertures.
We also observe a general increase in the depinning threshold
with increasing vortex density, 
which is opposite from the normal trend for vortices in two-dimensional
pinning arrays where the vortices are directly trapped by pinning sites. 
In the funnel geometry, the
pinning is a result of the repulsive vortex-vortex interaction forces,
and as the vortex density increases it becomes more difficult for 
the vortices to overcome these repulsive forces and 
flow through the funnel tip.
We also find a rich 
variety of dynamical phases, 
including ordered elastic and ordered plastic phases 
where the vortices follow fixed trajectories 
and disordered phases where vortices mix chaotically. 
The phases generally organize in such a way 
that only one vortex passes through the tip of the funnel at a time.
Due to this constraint, the average velocity of an individual 
vortex decreases with
increasing field such that the sum of the velocities of all of the vortices
at fixed drive remains close to constant for increasing field rather
than increasing with increasing field.  
This behavior is similar to the response of grains in
an hourglass.
Transitions between the different dynamical phases
appear as jumps or cusps in the 
velocity-force curves, and there are even regimes where the
average vortex motion decreases with increasing drive, 
producing a negative differential conductivity. 
At higher fields, moving states can form in which
a single vortex remains immobile at the center of a funnel 
while other vortices flow around it. 
In general, ordered flow phases occur at the matching fields, 
while at non-matching 
fields the flow is disordered for at least 
some regime of driving forces.  The ordered phases
are associated with sharp narrow band or washboard 
velocity noise signals while the disordered phases 
have $1/f^\alpha$ velocity noise spectra.  
Our results should also be generalizable 
to other systems of repulsively interacting particles
moving through a funnel geometry, such as colloids or Wigner crystals.     

We thank B. Plourde for helpful discussions. 
This work was carried out under the auspices of the 
NNSA of the 
U.S. DoE
at 
LANL
under Contract No.
DE-AC52-06NA25396.

\end{document}